\def\bey{\begin{eqnarray}}
\def\eey{\end{eqnarray}}
\def\be{\begin{equation}}
\def\ee{\end{equation}}
\def\ba{\begin{array}}
\def\ea{\end{array}}

\def\pp{\partial}
\def\pp{\partial}

\bigskip
\documentclass[prc,twocolumn,floatfix,groupedaddress,nofootinbib,showpacs,preprintnumbers,
amsmath,amssymb,amsfonts,superscriptaddress,widetable] {revtex4}
\usepackage{bm}
\usepackage{mathrsfs}
\usepackage{amssymb}
\usepackage{amsmath}
\usepackage{graphicx}
\usepackage{array}
\usepackage[usenames,x11names]{xcolor}
\usepackage[colorlinks,  linkcolor=red, anchorcolor=green, citecolor=blue]{hyperref}

\begin{document}
\preprint{ }
\title{Small radii of neutron stars as an indication of novel in-medium effects}
\bigskip
\author{Wei-Zhou Jiang}\email{wzjiang@seu.edu.cn}
\affiliation{Department of Physics, Southeast University,
             Nanjing 211189, China}
\affiliation{Department of Physics and Astronomy, Texas A\&M
               University-Commerce, Commerce, TX 75429, USA}
\author{Bao-An Li}\email{bao-an.li@tamuc.edu}
\affiliation{Department of Physics and Astronomy, Texas A\&M
               University-Commerce, Commerce, TX 75429, USA}
\author{F.~J. Fattoyev}\email{ffattoye@indiana.edu}
\affiliation{Department of Physics and Astronomy, Texas A\&M
               University-Commerce, Commerce, TX 75429, USA}
\affiliation{Center for Exploration of Energy and Matter, Indiana
               University, Bloomington, IN 47408, USA}

\date{\today}
\bigskip

\begin{abstract}
At present, neutron star radii from both observations and model
predictions remain very uncertain. Whereas different models can
predict a wide range of neutron star radii, it is not possible for
most models to predict radii that are smaller than about 10 km, thus
if such small radii are established in the future they will be very
difficult to reconcile with model estimates. By invoking a new term
in the equation of state that enhances the energy density, but
leaves the pressure unchanged we simulate the current uncertainty in
the neutron star radii. This new term can be possibly due to the
exchange of the weakly interacting light U-boson with appropriate
in-medium parameters, which does not compromise the success of the
conventional nuclear models. The validity of this new scheme will be
tested eventually by more precise measurements of neutron star
radii.
\end{abstract}

 \maketitle

Neutron star (NS) is a unique place to test fundamental forces at
the extremes of matter density, gravity and magnetic fields.
Unfortunately, uncertainties in both the Equation of State (EOS) of
super-dense nuclear matter and the strong-field gravity strongly
interplay with each other in determining observational properties of
neutron stars, for the latest review, see, {\sl e.g.,}  \cite{He15}.
For instance, in a simple version of modified gravity where the
non-Newtonian gravity exists, neutron stars could have very
different structures compared to predictions using Einstein's
General Relativity (GR) theory of gravity ~\cite{kr09,wen09,zh11}.
The radius of a neutron star is one of the most important
observables sensitive to the underlying nuclear EOS and gravity
theories used. Currently, within GR the radius of a canonical NS has
been predicted to be roughly from 11 to 15
km~\cite{wr88,la01,la07,li06,ji07} depending on the EOS used.
Provided the third family of compact stars known as strange stars
exist, their radii could be as small as 7 or 8
km~\cite{ka81,gl00,sh02}, although these models normally predict
star masses much smaller than the masses of observed massive neutron
stars. Thus, the measurement of NS radii plays a very important role
in resolving several issues in fundamental physics. Unfortunately,
the extraction of NS radii from observations still suffers from
large  systematic uncertainties~\cite{mil13} involved in the
distance measurements and theoretical analyses of the light
spectrum~\cite{la01,ha01,zh07,su11}. Consequently, a wide range of
the radius with the mass around 1.4 $M_\odot$ has been
reported~\cite{su11,st10,oz10,guv13,gu13,la13,st13,he14,gu14}. In
particular, using the thermal spectra from quiescent low-mass X-ray
binaries (qLMXBs) Guillot and collaborators extracted NS radii of
$R_{\rm NS} = 9.4\pm1.2$ km~\cite{gu14}. Another recent
comprehensive study of spectroscopic radius measurements suggest
that for a 1.5 $M_\odot$ NS the extracted radii are
$10.8^{+0.5}_{-0.4}$ km~\cite{oz15}. It is important to note that at
the moment no consensus has been reached yet on the extracted NS
radii. For instance,  Bogdanov found a 3-$\sigma$ lower limit of
11.1 km on the radius of the PSR J0437-4715~\cite{bog13}, and
Poutanen {\sl et al.} got a lower limit of 13 km for 4U
1608-52~\cite{pou14}. Whether the radii of canonical neutron stars
can be as low as 10 km have been an interesting topic of hot debate
during the last few years.

Whereas the situation has been significantly improved over the last
few years, the systematic errors have been hindering severely the
accurate determination of the NS radii from astrophysical
observations. However, if the existence of NSs with small radii is
firmly established, they would pose a severe challenge to the
current models of the nuclear EOSs. While it is not very difficult
to satisfy the maximum mass constraint, to our best knowledge,  no
nuclear models available with best-fit parameters to date can reproduce
the small NS radius constraint. There are only  a few microscopic
models or approaches that--- either disregard some of the nuclear
physics constraints~\cite{wr88} or adjust the high-density part of
the EOS using various polytropes to match the whole density profile
obtained from observation~\cite{heb13}---can account for  both
constraints. From the nuclear physics standpoint, the small NS
radius requires certain softness of the EOS of the
isospin-asymmetric nuclear matter. As one of the basic blocks of the
EOS of asymmetric matter, nuclear symmetry energy around 1-2 times
the saturation density of nuclear matter plays a dominating role in
determining the radii of neutron stars \cite{la01}. A softening of
the symmetry energy can lead to an appreciable decrease of the NS
radius~\cite{la07}. However, with the maximum NS mass held
approximately at a constant, most non-relativistic and relativistic
models that are facilitated with soft symmetry energies can only
bring the radii of canonical NSs down to about 12-13
km~\cite{li06,ji07,st10,la13}. Moreover, it becomes very difficult
to further reduce the NS radius by further softening the symmetry
energy. In particular, one would then encounter the stability
problem in the NS matter EOS when the symmetry energy becomes too
soft~\cite{wen09}. To further reduce the NS radius, one could
imagine to reduce the pressure of the isospin-symmetric part of the
EOS in the intermediate density region. But the space in so doing is
actually limited by the saturation properties of nuclear matter and
the constraint on the EOS of dense nuclear matter extracted from
studying nuclear collective flow in high energy heavy-ion
reactions~\cite{da02}. Furthermore, as we can see from the empirical
relation $Rp^{-1/4}(\rho_B)\approx ~C(\rho_B)$ between the NS radii
$R$ and the pressure $p(\rho_B)$ with $C(\rho_B)$ being a constant
at a given baryon density $\rho_B$~\cite{la07}, such a reduction is
also rather inefficient, since the isospin-symmetric EOS
contribution to the total NS matter pressure  $p(\rho_B)$ is
relatively small in the relevant density region, where this
empirical relation holds. Moreover, even if the significant
reduction were allowed for the total pressure, the limited decrease
of the NS radius would be at the cost of a large reduction of the NS
maximum mass, because the significantly reduced pressure needs the
corresponding reduction of the NS mass to balance the gravity
therein, and  also because the NS mass is the total energy
integrated in a nutshell with the reduced radius. A significant
reduction in the maximum mass is certainly disfavored by the
recently discovered massive neutron stars of $2 M_{\odot}$ that
require a stiff EOS~\cite{de10,an13}.

Facing currently with this severe theoretical issue, we explore in
this work a new possibility to soften the EOS: adding a new term in
the EOS that enhances the energy density while keeping the pressure
unchanged. In so doing, the enhancement of the energy density
requires some shrinkage in NS radius without reducing significantly
the NS maximum mass. Since the pressure is
$p=\rho_B^2\pp(\epsilon/\rho_B)/\pp\rho_B$, to keep the pressure
unchanged  we modify the energy density $\epsilon$ by
\begin{equation}\label{eq:eos1}
\epsilon=\epsilon_0+C_L \rho_B,
\end{equation}
where $\epsilon_0$ is the base energy density given by any model.
The second term {\sl linear in density} is an addition to the base
EOS with $C_L$ being \textsl{the amending coefficient}. As we shall
explain, this modification to the EOS of isospin-asymmetric nuclear
matter can be realized by considering the interaction added by a
vector boson with appropriate in-medium parameters.

\begin{figure}[thb]
\smallskip
\includegraphics[width=8.5cm,angle=0]{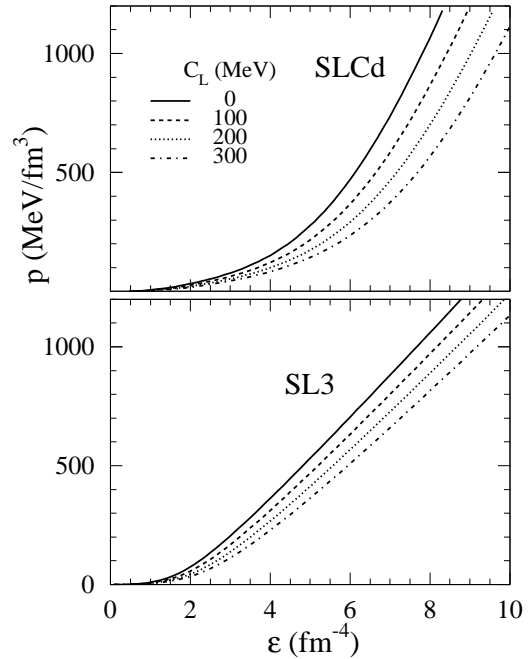}
\caption{The relation between the matter pressure and the energy
density  with  RMF models SLCd and  SL3. Note that the displacement
of energy density with a given $C_L$ is the same at a given density
but not at a constant pressure.\label{feos}}
\end{figure}

In principle, the modification given in Eq.~(\ref{eq:eos1}) could be
added to any nuclear model. In this work, we just demonstrate the
effects using several typical relativistic mean-field (RMF) models.
The RMF models under consideration include the SLC,
SLCd~\cite{ji07b}, SL3~\cite{ji07} and NL3040~\cite{pi04}. The SL3
and the NL3040 have similarly stiff EOSs at high density and both
give large NS maximum masses of more than 2.6$M_\odot$, while the
SLC and the SLCd feature the same EOS of symmetric matter within the
constraints obtained from analyzing the collective flow in
relativistic heavy-ion collisions~\cite{da02}. The only difference
between the SLC and the SLCd is that the latter has a softer
symmetry energy. In a similar fashion, the NL3040 was also built
from the original NL3 to feature a softer symmetry energy. The
stiffness of the symmetry energy is normally measured by its density
slope at the saturation density of nuclear matter $\rho_0$,
$L=3\rho_0\left(\pp E_{sym}(\rho_B)/\pp \rho_B\right)_{\rho_0}$. The
value of $L$ for the SLC and SL3  is 92.3 and 97.1 MeV, while it is
61.5 and 45.0 MeV for the SLCd and NL3040, respectively.  For a
comparison, it is interesting to note that currently the most
probable value of $L$ is in the range of $40\lesssim L \lesssim 70$
MeV according to recent analyses of various terrestrial experiments
and astrophysical observations, see, e.g.,
Refs.~\cite{ts12,ne13,ne09,st12,we12,jim13,ba13} and Ref.
\cite{EPJA} for a comprehensive review. Thus, the SLC and SL3 are
obviously too stiff while the SLCd and NL3040 are consistent with
the existing constraints in terms of their $L$ values. Nevertheless,
they are all appropriate for the purposes of this study.

Shown in Fig.~\ref{feos} are two examples of the EOS, i.e., pressure
versus energy density, with the SLCd and SL3 parameter sets. It is
seen that at a constant pressure, the amending term can soften the
EOS considerably, i.e., reducing the slope of the pressure with
respect to the energy density, especially with the SLCd. However,
the relative effect of the amending term goes down with the
increasing density because it is just linear in density while the
EOS of usual nuclear models evolves generally with the density
squared. We emphasize that the EOS softening scheme considered here
is quite different from the usual  mechanisms mentioned in the
introduction. In particular, typical phase transitions to matter
with new degrees of freedom normally reduce the maximum mass
dramatically, but often keep the NS radius more or less the same
because the phase transitions usually occur in the small inner core
of NSs. Of course, exceptions may exist when the new degrees of
freedom, such as the $\Delta$ resonances, can emerge at a relatively
low density~\cite{cai15}. It is interesting to note that by using
Lindblom's inversion algorithm~\cite{lind92} Chen and Piekarewicz
were recently able to obtain a softened EOS from the given small NS
radii~\cite{chen15}.

\begin{figure}[thb]
\smallskip
\includegraphics[width=8.5cm,angle=0]{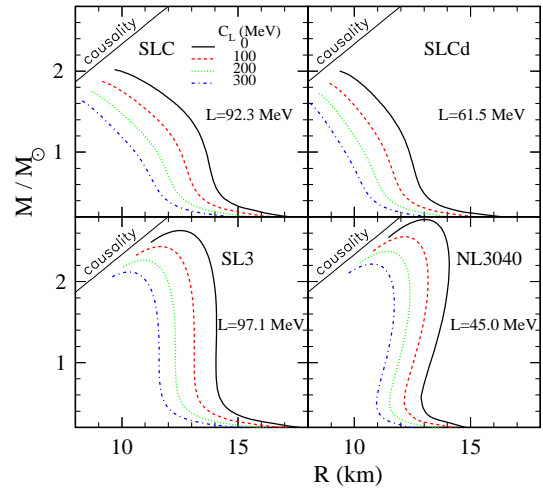}
\caption{(Color online) The mass-radius trajectories of neutron stars
with RMF models: SLC, SLCd, SL3 and NL3040. In each panel, the
amending coefficient $C_L$ is taken to be the values 0, 100, 200, 300
MeV, respectively. The slope parameter of the symmetry energy is also labeled in each panel.\label{msr1}}
\end{figure}

We now examine effects of the amending term on the radii of neutron
stars. As in Ref.~\cite{gu13}, here we consider the simplest model
of neutron stars consisting of just neutrons, protons and electrons.
Shown in Fig.~\ref{msr1} are the mass-radius (MR) trajectories of
neutron stars with the amending term within various RMF models. The
amending coefficient is exemplified as $C_L= 100$, $200$ and $300$
MeV, and the results with original models are displayed with
$C_L=0$. Comparing results of the original models,  we see that the
softening of the symmetry energy may reduce the NS radius by as
large as 1.5 km for a canonical NS when $L$ is reduced from 97.1 to
45 MeV. The space for further reducing the slope parameter $L$ is
small, and in fact, the further reduction in $L$ has a very limited
effect in decreasing the NS radius. Moreover, it is seen that even
with the significant softening of the symmetry energy within the
original models the NS radius is still far above $9.4\pm 1.2$ km
extracted by Guillot {\sl et al.} ~\cite{gu14}. If such small radii
are established, it is then interesting to see that the amending
term can indeed further reduce the NS radius. Obviously,  the role
of the amending term is similar in all models: the larger the
amending coefficient is, the more is the reduction of the NS radius.
With the same amending coefficient in different models, the shifted
magnitude is also similar. Typically, the amending coefficient,
varying up to 300 MeV, can cause a reduction of about 3 km in the NS
radius. With the amending coefficient of $C_L=300$ MeV, we see that
the MR trajectories with the SLC and SLCd fall into the regime
extracted by Guillot {\sl et al.}~\cite{gu13}.  We see from
Fig.~\ref{msr1} that the NS maximum mass is still not reached in the
SLC and SLCd models before reaching the causality boundary, because
the allowed nucleon density has a maximum in the construction of
such models to meet the chiral limit~\cite{ji07}. The removal of
such a limiting density may bring the NS maximum mass closer to the
causality limit. With the larger amending coefficient, the MR
trajectories for the SL3 and NL3040 can be very close to the upper
margin of the extracted regime. Notably, we see that the MR
trajectories are not clearly away from the causality constraint,
though the amending coefficient causes the decrease of the NS
maximum mass. Here, the moderate reduction of the NS maximum mass is
just because of the softening of the EOS, namely no excess of
pressure can resist the additional gravity arising from the increase
of the energy density.

\begin{figure}[thb]
\smallskip
\includegraphics[width=8.5cm,angle=0]{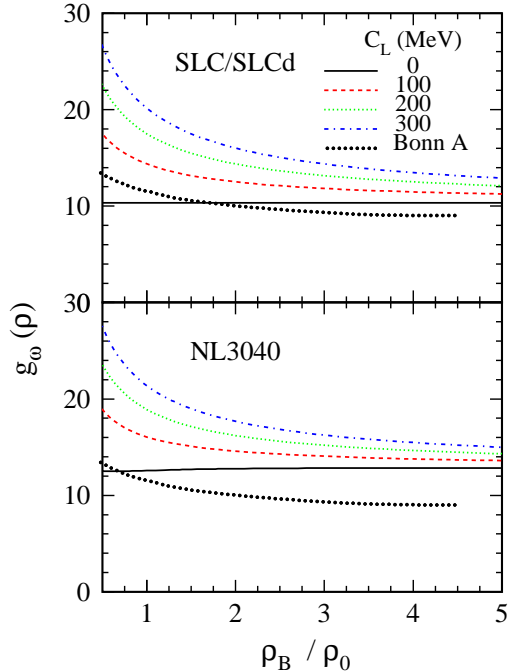}
\caption{(Color online) The density-dependent vector coupling
constant  incorporated from the amending term in RMF models: SLC/SLCd
and NL3040. The density-dependent coupling constant obtained from the
DB potential of Bonn A is also given for comparison. \label{figcpl}}
\end{figure}

In the RMF framework, the amending term in Eq.~(\ref{eq:eos1}) can
be understood as a specific in-medium effect. Similar to the
analysis in Refs.~\cite{br92,fr93,ma94}, the amending term,
incorporated into the vector potential, leads to the
density-dependent coupling constant of the vector ($\omega$) meson
\begin{equation}
g_\omega^2(\rho_B)=(g_\omega\omega +2C_L)m_\omega^2/\rho_B.
\end{equation}
This relation indicates that the larger $C_L$ is responsible for the
stronger density dependence of the coupling constant, and with the
increase of density, the in-medium effect decreases with the growing
$\omega$. Fig.~\ref{figcpl} shows the density-dependent coupling
constant for the models considered. For a comparison, the
$g_\omega(\rho_B)$ obtained from the Dirac-Brueckner (DB) potential
of Bonn A is also shown in the figure. It is seen that the density
dependence, similar to the one from the DB potential, is needed to
produce a significant reduction of the NS radius. We can infer,
indeed, that the density dependence here is stronger than that from
the DB potential, because the latter owns a partial cancelation of
the density dependencies between the scalar and vector
potentials~\cite{br92,fr93,ma94}. In addition, we see that the
in-medium effect of all cases tend to vanish at high densities. This
means that the decrease of the NS radius is dominated by  the
modification of the amending term to the low-density EOS. For the
dropping of the NS maximum mass,  it can then hopefully be cured by
modifying the high-density component of the EOS of nuclear matter.

The DB potential is obtained by solving the two-body correlations.
One may expect that the many-body rather than just two-body
correlations may generate more in-medium effects~\cite{wr88}. It
should, however, be pointed out that the arbitrarily strong density
dependence usually lacks observable grounds. In the present RMF
models, one should evade the violation of the low-energy constraints
in finite nuclei from the in-medium effect induced by the amending
term. One possible solution is to resort to the exchange of the very
light and weakly interacting boson that is undetected to date. A
favorite candidate is the U-boson, which is actually a particle
beyond the standard model~\cite{fa80,fa86}. The light U-boson, first
proposed by Fayet~\cite{fa80}, might be regarded as the mediator of
the putative fifth force~\cite{fa86,fi99,ad03}. Recently, the
possibility of the MeV dark matter with the light U-boson mediation
was considered to account for the bright 511 keV $\gamma$ emission
of positron annihilations from  the galactic
bulge~\cite{bo04,boe04,bor06,zhu07,fa07,je03}, despite that there
are a number of conventional astrophysical sources for positron
annihilations, see Ref.~\cite{pra11} and references therein.   In
the past, the significant effects of the U-boson in neutron stars
were predicted~\cite{kr09,wen09,zh11}, albeit the boson should
couple to baryons very weakly.  The modification of the EOS due to the
U-boson in the mean-field approximation simply reads
\begin{equation}
\epsilon_{UB}=\frac{1}{2}\left(\frac{g_u}{m_u}\right)^2\rho_B^2,
\end{equation}
with $g_u/m_u$ being the ratio of the U-boson coupling constant and
its mass.

By assuming a density-dependent U-boson mass of
$m_u^2=g_u^2\rho_B/2C_L$,  we realize the linear density dependence of the
amending energy density. If the U-boson has a very light bare mass of
$m_{u0}$ in free space, its in-medium mass can be given as
\begin{equation}\label{equbn}
m_u^2=g_u^2\rho_B/2C_L+m_{u0}^{2}.
\end{equation}
In general, the in-medium effect of the boson parameters is
associated with the contribution of the intermediate states. In the
RMF theory, the in-medium effect is usually attributed to the
nonlinearity of the meson self-interactions~\cite{bo77}. The
in-medium effect for the boson may also be realized by carefully
choosing the nonlinear self-interacting terms. We leave this problem
for a future study.

\begin{figure}[thb]
\smallskip
\includegraphics[width=7.5cm,angle=0]{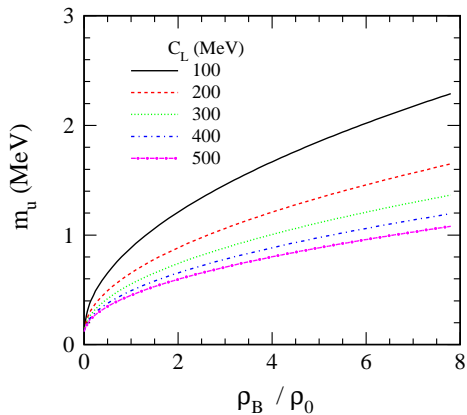}
\caption{(Color online) The in-medium mass of the U-boson with
various $C_L$ as a function of density. Here, $g_u=0.01$ and
$m_{u0}=0.1$MeV. \label{umss}}
\end{figure}

Regarding the possible violation of the low-energy nuclear
constraints for finite nuclei, one can avoid it by limiting the weak
interaction strength of the U-boson. For instance, if $g_u$ is 0.01,
the interaction strength is about two orders of magnitude weaker
than the electromagnetic interaction, and such weak interaction is
not able to affect properties of finite nuclei. In this sense, the
present scheme to invoke the U-boson does not compromise the success
of optional nuclear models. The interesting question is what
behavior of the U-boson mass will allow such a weak interaction
strength. Shown in Fig.~\ref{umss} is the density-dependent mass of
the U-boson as a function of density, for $g_u=0.01$. Here, as an
example, the small bare boson mass is taken to be as 0.1 MeV, which
is just a free parameter.  We see that for the given coefficients
$C_L$, the in-medium mass is just within a few MeV. With the
increase of the $C_L$, the in-medium mass of the U-boson  decreases.
For $C_L=300$ MeV, the U-boson mass is close to 1 MeV, which is
consistent with that considered in Ref.~\cite{zh11}.

We should say that the role of the vector boson in softening the EOS
eventually is rather pragmatic, albeit putatively accounting for  small NS radii.
Usually, the vector boson is a source of
the pressure, while here it gives zero contribution. From the point
of view of nuclear many-body approaches, the repulsion of the vector
meson can be softened in the intermediate density region due to the
correlation effect from the intermediate-state contributions, and
such a softening can be simulated in the RMF theory by invoking the
nonlinear self-interacting terms. At high densities where the
intermediate-state contribution is small due to the Pauli blocking
the reduction of the in-medium effect of the vector boson will then
recover the repulsion which may enable a stiffer EOS at high
densities. In this sense, if we  use a more complicated
density-dependent term in Eq.(\ref{eq:eos1}) in the high density
region, the corresponding U-boson can then stiffen the high-density
EOS and the U-boson mass would be also much closer to a constant at
high densities.  While we use a simple amending term in
Eq.(\ref{eq:eos1}) to demonstrate the reduction of the NS radius, it
should not  represent that the corresponding vector U-boson would
simply soften the EOS.

The NS radius can attain larger reduction by further increasing the
amending coefficient. In our analysis here we have invoked the
special amending term that does not affect the pressure. Generally,
other forms could also be optional, as long as they result in the
reduction of the NS radius. A favorable form should reduce the
low-density pressure and satisfy all constraints on the EOS of
nuclear matter at saturation density. To meet these demands, one can
consider the light scalar boson. Note that the scalar boson, like
the scalar meson, can also modify the mass of baryons. We have
checked that the inclusion of such a light scalar boson can again
reduce the NS radius by 0.5-1.5 km, depending on the model used. To
keep the pressure positive definite, the coupling constant of the
scalar boson should be much weaker (e.g., $g_s=0.005$) than that of
the vector boson, and the NS maximum mass is little changed by the
scalar boson.

In summary, we have demonstrated that the large uncertainty  of the
NS radius can be simulated by introducing the amending term that
enhances the energy density, but leaves the pressure unchanged. The
distinct effect of the amending term on the reduction of the NS
radius is rather universal to any nuclear EOS. The new term is
linear in density, and the incorporation into the relativistic
mean-field potential produces the strong in-medium effect on the
vector coupling constant at low densities.  The  strong in-medium
effect can be responsible for the significant reduction of neutron
star radii, and may provide an explanation for the small NS radius
if it is firmly established. We interpret this novel in-medium
effect by the possible exchange of a putative weakly interacting
U-boson. By requiring that the low-energy constraints in finite
nuclei must not be violated, we find interestingly that the
in-medium U-boson mass is just below a few MeV with the interaction
strength being two orders of magnitude weaker than the
electromagnetic interaction. Finally, the validity of the scheme
presented here will be tested eventually by the more precise
measurements of the NS radii.

\section*{Acknowledgement}
We would like to thank W.G. Newton for helpful discussions.
The work was supported in part by the National Natural Science
Foundation of China under Grant No. 11275048 and 11320101004, the China Jiangsu
Provincial Natural Science Foundation under Grant No. BK20131286,
the US National Science Foundation under grants No. PHY-1068022 and PHY-1205019,
the U.S. Department of Energy's Office of Science under Award Number DE-SC0013702,  the CUSTIPEN (China-U.S. Theory Institute for Physics with Exotic Nuclei) under DOE
grant number DE-FG02-13ER42025, and DOE grants DE-FG02-87ER40365 (Indiana University) and DE-SC0008808 (NUCLEI SciDAC Collaboration).

\end{document}